\documentclass[11pt]{article}

\usepackage[preprint]{acl}

\usepackage{times}

\usepackage{latexsym}
\usepackage[T1]{fontenc}
\usepackage[utf8]{inputenc}
\usepackage{microtype}
\usepackage{graphicx}
\usepackage{booktabs}
\usepackage{array}
\usepackage{amsmath}

\title{Post-Generation Verification Dominates Retrieval Optimization:\\
A $2^4$ Factorial Ablation of RAG Pipeline Features}

\newif\ifanon \anonfalse
\ifanon
  \author{Anonymous ACL submission}
\else
  \author{Ng S.\ T.\ Chong \\
    Campus Computing Centre, United Nations University \\
    Tokyo, Japan \\
    \texttt{ngstc@unu.edu}}
\fi

\newif\iftocpub \tocpubtrue

\begin{document}
\maketitle

\begin{abstract}
Modern RAG pipelines stack many enhancement features, but these
features are typically validated in isolation, leaving their
interactions unmeasured. We run a $2^4$ full factorial ablation of four
pipeline features: section expansion (SE), agentic search (AS),
completeness check (CC), and table-of-contents-guided retrieval (ToC).
The design crosses 16 configurations, 24 queries spanning eight
interaction types, and two cloud-class models (768 designed runs, 759 completed and scored) on
five public documents (78--492 pages), and every answer is scored
against a verified reference. On this corpus and task, post-generation
verification dominates: CC is the strongest
feature ($d=+0.48$, $p<0.001$), improving accuracy, completeness, and
usefulness simultaneously, and CC alone (4.31/5) is ahead of every
configuration without it, including the three-feature SE+AS+ToC (4.11).
ToC yields a significant gain at zero additional LLM calls ($d=+0.22$); AS shows a
small, unstable effect that does not survive correction for multiple
comparisons, helping some queries and harming others; SE is neutral.
These main effects hold when every response is re-scored by two
alternate LLM judges. The highest-quality configuration roughly doubles
baseline latency,
producing a genuine quality--latency Pareto frontier of six
configurations. Feature utility is strongly query-type dependent (CC's
completeness effect reaches $d=+0.83$ on completeness-demanding queries), so
single-query-type evaluations systematically mis-rank features. We
conclude that verifying answers matters more than optimizing retrieval,
and that factorial designs with diverse query types are necessary to
evaluate RAG features.
\end{abstract}

\section{Introduction}

RAG pipelines have grown from simple retrieve-then-generate
architectures into multi-stage systems with numerous enhancement
features \citep{gao2024}. Each feature (context expansion, query
decomposition \citep{ma2023,trivedi2023}, answer verification
\citep{asai2024,shao2023}, structure-aware retrieval
\citep{jeong2025}) is typically justified by a targeted
improvement on a specific failure mode. Yet features are deployed
simultaneously, and their interactions are rarely studied.

The standard methodology of testing each feature individually against
the baseline cannot capture interaction effects, diminishing returns, or
structural dependencies between features. A factorial design, in which
every combination of feature states is tested, is required to estimate
main effects cleanly and to detect interactions. Factorial ablations of
RAG pipeline features remain rare, partly because the number of
conditions grows exponentially.

We conduct a $2^4$ full factorial ablation of four features: section
expansion (SE), agentic search (AS), completeness check (CC), and
table-of-contents-guided retrieval (ToC). The ablation crosses 16
configurations, 24 queries spanning eight interaction types, and two
cloud-class models, totaling 768 designed runs (759 completed and scored). The corpus comprises five publicly available
documents (78--492 pages) covering data protection, human resources, AI
regulation, procurement, and cybersecurity. All four features operate
independently, with no architectural dependencies between them, ensuring
clean main-effect estimation.

The results show a clear hierarchy: CC dominates all quality dimensions,
ToC provides a significant improvement at zero additional LLM calls, AS helps on some query
types but hurts on others, and SE is neutral on this corpus. The
all-features configuration is highest-quality but \emph{not}
lowest-latency, revealing a genuine quality--latency Pareto tradeoff with
six frontier configurations. Crucially, feature utility is strongly
moderated by query type, with aggregate effects masking substantial
targeted variation.

\paragraph{Contributions.}
(1) A $2^4$ factorial ablation on a corpus of public documents, to our
knowledge the first RAG feature ablation at this scale (16 configurations
$\times$ 24 queries $\times$ 2 models $=$ 768 runs, 759 scored) in which all
features operate independently.
(2) Evidence that post-generation verification (CC) matters more than
retrieval-stage optimization, with CC alone (4.31/5) ahead of every
non-CC configuration (a descriptive comparison of configuration means;
see Table~\ref{tab:maineffects} for the formally tested main effect).
(3) Page-level retrieval precision and recall for ToC, the one feature
whose evidence carries page provenance, measured against an
independently cross-checked ground truth.
(4) A feature-interaction query taxonomy that enables targeted evaluation
of each pipeline stage and its combinations.
(5) Exploratory evidence that feature utility is strongly query-type
dependent, with aggregate effects masking large targeted differences.

\section{Related Work}

\paragraph{RAG architectures and feature engineering.}
\citet{lewis2020} introduced RAG as a general-purpose recipe combining
parametric and non-parametric memory. Subsequent work extended
retrieve-then-generate with iterative retrieval \citep{shao2023},
self-reflective generation \citep{asai2024}, and active retrieval that
interleaves reasoning with search \citep{jiang2023}. \citet{gao2024}
survey the modular RAG landscape, categorizing enhancements into
pre-retrieval, retrieval, and post-retrieval stages; our four features
map onto this taxonomy, with CC extending it to a post-generation stage. Most proposed features are evaluated in isolation
against the baseline; factorial studies that quantify interaction effects
remain rare.

\paragraph{Query decomposition.}
Decomposing complex queries into sub-queries is central to multi-hop
question answering. \citet{trivedi2023} interleave chain-of-thought
reasoning with retrieval; \citet{ma2023} show that LLM-based query
rewriting improves retrieval for ambiguous queries. Our AS feature
implements LLM-based decomposition in this spirit.

\paragraph{Structure-aware retrieval.}
Document structure (headings, tables of contents, section hierarchies)
provides signals that complement dense similarity. Recursive language
models overcome context limits through hierarchical reasoning
\citep{zhang2025}. DocsRay \citep{jeong2025} generates a pseudo table of
contents via multi-phase LLM processing and interpolates heading and
content scores for within-section retrieval. Our ToC feature differs in
three ways: headings are extracted without LLM calls (native PDF
bookmarks or layout-inferred hierarchy), matched sections are loaded as
full pages rather than sub-chunks, and heading and chunk indexes operate
as independent parallel channels rather than blended scores. These
choices remove LLM indexing cost and the blending coefficient that
couples the two retrieval signals.\ifanon\else\iftocpub\ A fuller
treatment of the ToC method and its factorial evaluation appears in
\citet{tocpaper}.\fi\fi

\paragraph{Completeness verification.}
Post-generation loops that check whether an answer covers all query
aspects have been proposed in several forms: \citet{asai2024} use
self-reflection to decide when to re-retrieve, and \citet{shao2023}
iterate retrieval--generation cycles. Our CC feature implements a
loop-back mechanism using \emph{facet decomposition}: it decomposes the
question into distinct information facets, checks each against the
answer, and triggers re-retrieval for missing facets. Prompt design is
critical: holistic ``is this complete?'' assessments produce high
false-negative rates, while facet-decomposition prompts substantially
reduce them.

\paragraph{LLM-as-judge and factorial designs.}
Using strong LLMs to evaluate LLM outputs is now standard
\citep{zheng2023}. We use exclusively model-answer anchored evaluation,
scoring each response against a verified reference to address the
reference-free evaluation bias raised by \citet{shankar2024}. Full
factorial designs are common in experimental psychology but rare in NLP
system evaluation; \citet{dodge2019} vary random seeds (weight
initialization and data order) rather than architectural features when
fine-tuning, but we are unaware of prior factorial ablations of RAG
pipeline features at this scale.

\section{System and Method}

\subsection{Pipeline and Features}

The system is a multi-stage RAG pipeline built on FastAPI with
PostgreSQL/pgvector for hybrid retrieval (cosine similarity + BM25). Each
query passes through domain classification, embedding, hybrid retrieval
(top-$k$ 8, fused from 32 vector and 32 BM25 candidates), optional feature
stages, answer generation, and optional verification. Table~\ref{tab:features} describes the four features under
test. AS's query decomposer and CC's facet checker run on a local review
model (llama3.1:8b); the two cloud models generate the answers
(Appendix~\ref{sec:settings}). All four operate independently, with no architectural dependencies
or guard conditions, ensuring a clean orthogonal factorial design.

\begin{table*}[t]
\centering
\small
\begin{tabular}{@{}llp{7.4cm}l@{}}
\toprule
\textbf{Feature} & \textbf{Stage} & \textbf{Mechanism} & \textbf{Cost} \\
\midrule
SE  & Retrieval & Retrieves sibling chunks sharing the same heading breadcrumb & $\sim$0 LLM calls \\
AS  & Pre-retrieval & An LLM decomposes each query into up to 4 sub-queries and retrieves for each (forced mode; the production heuristic gate was off) & 1 call $+ N$ retr. \\
CC  & Post-generation & Decomposes the question into facets, checks each against the answer, re-retrieves and regenerates for missing facets & 1--2 calls/hop \\
ToC & Retrieval & Matches the query against pre-embedded ToC entries, loads full page sections from a sidecar, bypassing the chunk layer & $\sim$0 LLM calls \\
\bottomrule
\end{tabular}
\caption{The four features under test: section expansion (SE), agentic
search (AS), completeness check (CC), and ToC-guided retrieval (ToC). SE
and ToC are retrieval-stage features that add zero additional LLM calls;
AS is a pre-retrieval LLM feature; CC is a post-generation verification
feature.}
\label{tab:features}
\end{table*}

\paragraph{CC prompt design.}
The completeness checker (1) decomposes the question into distinct
information facets, (2) checks each facet against the answer as
covered/missing/partial, and (3) checks the breadcrumb structure for
numbered-sequence gaps. The answer is marked incomplete if any facet is
missing or multiple are partial. This decomposed design is essential:
holistic prompts (``default to complete, only flag if a topic is not
addressed at all'') anchor on a lenient prior and produce high
false-negative rates.

\paragraph{ToC rationale.}
Chunk-based retrieval has two blind spots that neither a larger
\texttt{top\_k} nor SE can address. First, SE fetches sibling chunks
under the same parent heading but cannot reach \emph{cousin} sections
(sub-trees under a different branch of the hierarchy). Second, some
sections have no sub-headings, so chunk retrieval cannot distinguish
sub-topics within them. ToC addresses both at a different retrieval
granularity: at ingestion every document's headings are extracted and
each title is pre-embedded; at query time, cosine similarity against
these entries identifies relevant sections, and full page content is
loaded from a sidecar file, bypassing the chunk layer and preserving
intra-section context that chunking fragments.

\subsection{Corpus and Experimental Design}

The corpus comprises five publicly available documents of varying length
and structural complexity: GDPR \citep{gdpr2016} (78 pp., data
protection), the UN Staff Rules \citep{unstaffrules2023} (118 pp., human
resources), the EU AI Act \citep{euaiact2024} (144 pp., AI regulation),
the UNOPS Procurement Manual \citep{unopsprocurement2021} (195 pp.,
procurement), and NIST SP 800-53r5 \citep{nist80053r5} (492 pp.,
cybersecurity). They span shallow-to-deep heading hierarchies and dense
cross-referencing. The deployment's retrieval index also held a few
internal institutional documents, which shaped two reference answers
(\S\ref{sec:retrieval-robustness}).

We use a $2^4$ full factorial design (16 configurations) $\times$ 24
queries $\times$ 2 models $=$ 768 conditions. The two models are
DeepSeek-V4-Flash \citep{deepseekv4} and grok-4.3 \citep{grok43} (both
via Azure AI), chosen as cloud-class models of comparable capability to
test for model dependence without a model-tier confound. The 24 queries span eight interaction
types, each designed to exercise a specific feature mechanism or
combination (Table~\ref{tab:querytypes}). This taxonomy is itself a
contribution: it lets us read each feature's effect where the feature is
supposed to matter, rather than only in aggregate.

\begin{table}[t]
\centering
\small
\begin{tabular}{@{}llp{3.7cm}@{}}
\toprule
\textbf{Type} & \textbf{$n$} & \textbf{Target} \\
\midrule
baseline    & 3 & None (baseline sufficient) \\
se\_target  & 3 & SE \\
toc\_target & 4 & ToC \\
as\_target  & 3 & AS \\
cc\_target  & 3 & CC \\
toc\_cc     & 3 & ToC + CC \\
as\_toc     & 2 & AS + ToC \\
full\_stack & 3 & SE + AS + CC + ToC \\
\bottomrule
\end{tabular}
\caption{Query interaction taxonomy. Each type targets a specific
pipeline stage or feature combination.}
\label{tab:querytypes}
\end{table}

\subsection{Evaluation}

\paragraph{Model-answer anchored scoring.}
Each completed response is scored against a gold-standard reference
answer generated from the source documents (two exceptions,
\S\ref{sec:retrieval-robustness}) and manually verified for factual
accuracy. Each response receives \emph{accuracy} (factual correctness
against the reference), \emph{completeness} (fraction of the
reference's content covered), and \emph{usefulness} (practical
actionability) scores (1--5 Likert; full rubric in
Appendix~\ref{sec:repro}); composite quality is their unweighted mean. The
scorer is Claude Opus 4.6 \citep{claudeopus46}, called directly (not
through the RAG pipeline) to avoid retrieval contamination, at
temperature 0.

\paragraph{Judge robustness.}
\label{sec:judgerobust}
A single LLM judge invites an obvious objection: are these scores an
artifact of how one model interprets the rubric? We checked this five
ways. \emph{Two independent human raters} each blind-scored the same
stratified sample of 24 responses (one per query, spanning 13 of the 16
configurations and both answer models), without seeing the judge's
scores or each other's. \emph{Two independent LLM judges} re-scored all
758 scored responses on the original prompt: gpt-6-sol and Claude Opus
5.5. Both were chosen specifically because neither provider generated
any of the candidate answers under evaluation. DeepSeek-V4-Flash and
grok-4.3 are the two answer models, so using either of them (or a
sibling model from the same family, e.g.\ a larger DeepSeek variant) as
a judge would risk a self- or family-evaluation bias on exactly the
responses that model or its relative produced. gpt-6-sol provides a
cross-provider check (OpenAI, independent of both answer-model
families); Opus 5.5 provides a same-family, later-generation check
(does a newer Claude model agree with an older one on this exact
task?). Finally, the \emph{same} judge re-scored all 758 responses on a
reworded prompt that asks for the same three scores with different
phrasing and structure, to check sensitivity to prompt wording rather
than model identity. Each re-scoring run lost one response (a different
one each time) to a timeout or unparseable output, leaving 758.

Table~\ref{tab:judgerobust} summarizes composite-quality agreement with
Claude Opus 4.6 for each check. The result we did not expect: all four
alternates disagree with Opus 4.6 in the \emph{same} direction. Two
independent humans and two alternate LLM judges (one cross-provider, one
a later model from the same family) all score more strictly than
Opus 4.6, not more leniently and not inconsistently in either direction.
This is itself evidence of a \emph{convergent calibration offset} rather
than random noise, although the moderate correlations ($r=0.64$--$0.84$)
also show some genuine disagreement about individual responses. The two human raters also agree with
\emph{each other} at $r=0.61$ (bias $-0.53$), a similar magnitude to
their individual agreement with the judge ($r=0.69$ and $r=0.64$); this
indicates that part of the measured disagreement reflects genuine
subjectivity in 1--5 answer-quality judgments (two reasonable humans
scoring the same response do not perfectly agree with each other either)
rather than a defect specific to the LLM judge.

Because every feature comparison in this paper is differential (feature
ON minus OFF, computed with the same judge, prompt, and temperature
throughout), a roughly constant offset of this kind cancels in the
subtraction. It would only affect absolute statements, which we report
descriptively and do not rely on for the paper's comparative claims
(e.g., per-type difficulty in Table~\ref{tab:querytype} and the share of
low-scoring responses in \S\ref{sec:retrieval-robustness}): we never
claim that a 4.31/5 score means the pipeline is
``good'' in any absolute sense, only that it is higher than the
alternatives under the same scoring procedure. The reworded-prompt
check additionally shows the judge is highly stable to prompt wording
specifically ($r=0.95$, bias $+0.05$), isolating prompt sensitivity from
judge-identity effects: the disagreement with alternate judges/raters is
not explained by brittleness to how the question is phrased. Most
directly, recomputing the four main effects under each alternate scoring
reproduces the paper's pattern: CC ($d=+0.43$ to $+0.55$) and ToC
($d=+0.17$ to $+0.25$) remain significant after FDR correction under
gpt-6-sol, Opus 5.5, and the reworded prompt; SE stays null; and AS
reaches significance only under the reworded prompt. CC alone also stays
ahead of every configuration without CC under all three.

\begin{table}[t]
\centering
\small
\begin{tabular}{@{}lccc@{}}
\toprule
\textbf{Check} & \textbf{$n$} & \textbf{Bias} & \textbf{$r$} \\
\midrule
Human rater 1 (blind) & 24 & $-0.43$ & 0.69 \\
Human rater 2 (blind) & 24 & $-0.96$ & 0.64 \\
gpt-6-sol (cross-provider) & 758 & $-0.85$ & 0.80 \\
Opus 5.5 (newer, same family) & 758 & $-0.53$ & 0.84 \\
Reworded prompt (same judge) & 758 & $+0.05$ & 0.95 \\
\bottomrule
\end{tabular}
\caption{Judge-robustness checks on composite quality. Bias $=$
alternate $-$ Claude Opus 4.6 (the paper's judge); $r$ is the Pearson
correlation with Opus 4.6's scores across the checked responses. All
four alternate judges/raters agree with Opus 4.6 on relative ordering
($r\ge0.64$) while scoring more strictly in a consistent direction.
Per-dimension breakdowns are not computed; the two human raters scored a
stratified sample of 24 responses, one per query, covering 13 of the 16
configurations and both answer models, sampled independently at random
(seed 42) and blind to the judge's scores and to each other's.}
\label{tab:judgerobust}
\end{table}

\paragraph{Retrieval metrics.}
Page-level precision and recall compare retrieved page numbers against
labeled relevant pages. Because only ToC-sourced evidence encodes page
numbers in its chunk IDs, while regular chunks use hash-based IDs with no
page mapping, retrieval metrics are measurable only for ToC. This is a property of our
instrumentation, not evidence that the other features leave retrieval
unchanged: SE, for example, adds $\sim$3.9 chunks per query
(\S\ref{sec:retrieval-robustness}).

\paragraph{Statistics.}
Main effects use Welch's $t$-test; effect sizes are Cohen's $d$ with
5{,}000-resample bootstrap 95\% CIs (seed 42); the four main-effect
$p$-values are the paper's planned comparisons and are corrected
jointly with Benjamini--Hochberg FDR. Query-type subgroup tests and the
eleven factorial interaction contrasts are exploratory and reported
uncorrected. Full settings are in Appendix~\ref{sec:settings}.

\section{Results}

\subsection{Overall Quality}

Of 768 designed conditions, 760 completed (8 failures from transient API
errors), and 759 received valid judge scores (the scoring call failed
for one response). All quality statistics use these 759; latencies use
all 760. The judge-robustness re-scorings (Table~\ref{tab:judgerobust})
each lost one further response, a different one per run, leaving 758.
Mean composite quality is 4.04 (SD $=0.83$). The two models are
statistically indistinguishable (DeepSeek-V4-Flash $4.05\pm0.80$ vs.\
grok-4.3 $4.03\pm0.86$, $p=0.72$). Table~\ref{tab:configs} reports quality and
latency by configuration; the quality range across configurations is
0.75 points (3.60--4.35), indicating substantial differentiation. CC
alone reaches 4.31, above every configuration that lacks CC, including
the three-feature SE+AS+ToC (4.11).

\begin{table*}[t]
\centering
\footnotesize
\begin{tabular}{@{}lccccc@{}}
\toprule
\textbf{Config} & \textbf{Accuracy} & \textbf{Completeness} & \textbf{Usefulness} & \textbf{Quality} & \textbf{Latency} \\
\midrule
Baseline       & 3.77 & 3.30 & 3.74 & 3.60 & 78s \\
SE             & 3.77 & 3.35 & 3.73 & 3.62 & 80s \\
AS             & 3.77 & 3.53 & 4.06 & 3.79 & 115s \\
ToC            & 3.96 & 3.69 & 4.04 & 3.90 & 107s \\
CC             & 4.19 & 4.21 & 4.54 & \textbf{4.31} & 169s \\
SE+AS          & 3.81 & 3.53 & 3.98 & 3.77 & 82s \\
SE+ToC         & 4.00 & 3.69 & 4.00 & 3.90 & 80s \\
SE+CC          & 4.02 & 4.06 & 4.40 & 4.16 & 208s \\
AS+ToC         & 4.00 & 4.00 & 4.27 & 4.09 & 95s \\
AS+CC          & 4.02 & 3.92 & 4.19 & 4.04 & 238s \\
CC+ToC         & 4.20 & 4.17 & 4.50 & 4.29 & 206s \\
SE+AS+ToC      & 4.04 & 4.00 & 4.29 & 4.11 & 103s \\
SE+AS+CC       & 4.11 & 4.19 & 4.60 & 4.30 & 241s \\
SE+CC+ToC      & 4.09 & 3.87 & 4.30 & 4.09 & 211s \\
AS+CC+ToC      & 4.26 & 4.21 & 4.55 & 4.34 & 230s \\
SE+AS+CC+ToC   & 4.21 & 4.25 & 4.60 & \textbf{4.35} & 175s \\
\bottomrule
\end{tabular}
\caption{Composite quality (1--5) and latency by configuration, averaged
across both models. CC alone (4.31) exceeds every non-CC configuration.
The all-features configuration is highest-quality (4.35) but more than
doubles baseline latency. Each condition ran once; standard errors are
about 0.1 for quality and 4--21s for latency (see \S\ref{sec:discussion}
on latency gaps involving SE).}
\label{tab:configs}
\end{table*}

\subsection{Main Effects}
\label{sec:main-effects}

\begin{figure}[tbp]
\centering
\includegraphics[width=\columnwidth]{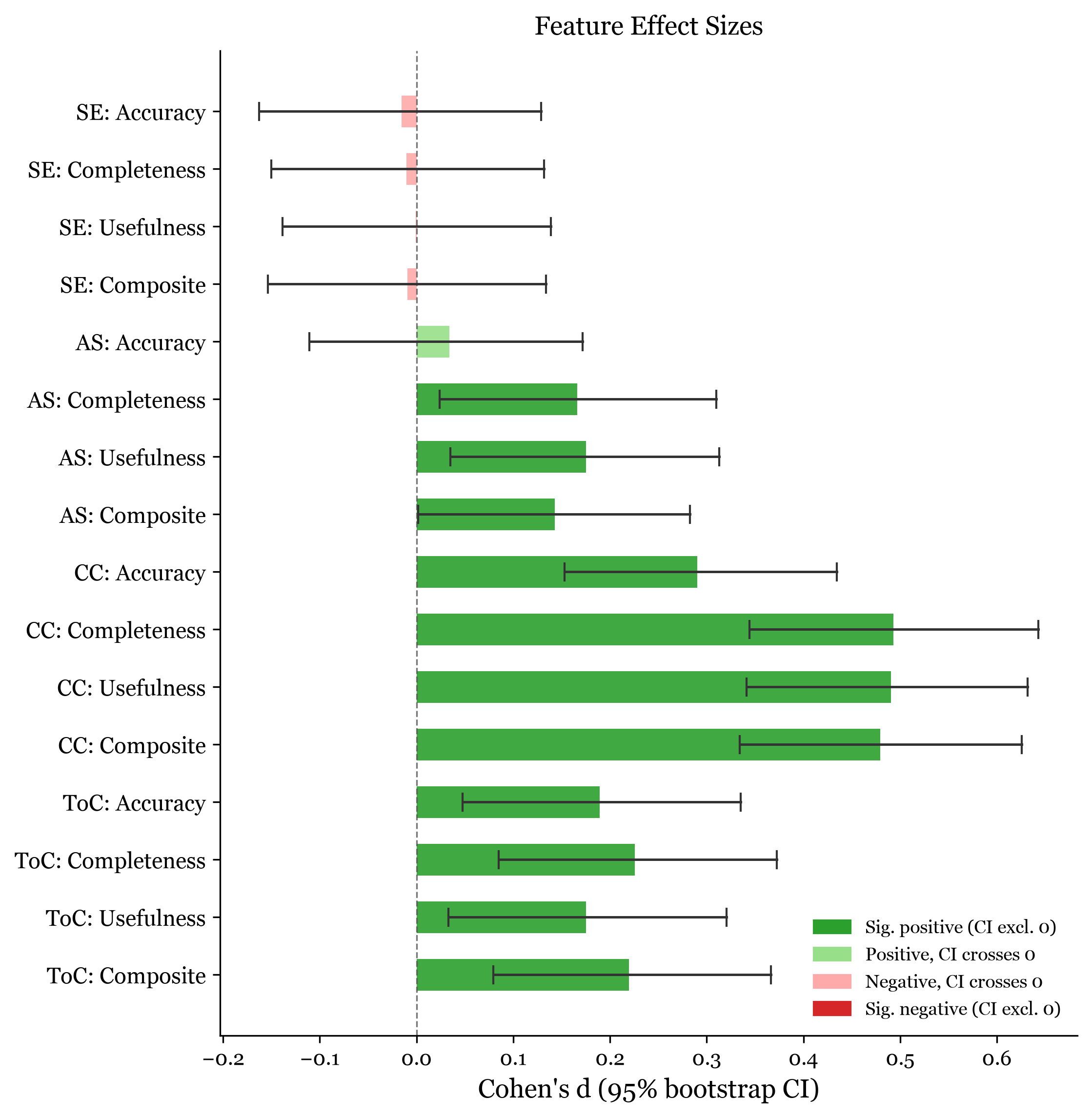}
\caption{Cohen's $d$ effect sizes for each feature across the four
quality dimensions, with 95\% bootstrap CIs (5{,}000 resamples). Dark
bars: significant positive (CI excludes 0); light bars: non-significant.
CC's intervals exclude zero on all dimensions; ToC excludes zero on all
three scoring dimensions; AS is significant on completeness and
usefulness, and its composite CI only narrowly excludes zero (raw
$p=0.050$; not significant after FDR correction,
Table~\ref{tab:maineffects}); SE is uniformly null.}
\label{fig:forest}
\end{figure}

Table~\ref{tab:maineffects} and Figure~\ref{fig:forest} report the main
effects. CC is the dominant feature ($d=+0.48$, $p<0.001$), significant
across all three dimensions: accuracy ($\Delta=+0.25$, $d=+0.29$),
completeness ($\Delta=+0.47$, $d=+0.49$), and usefulness ($\Delta=+0.44$,
$d=+0.49$). ToC shows a small effect ($d=+0.22$, raw
$p=0.003$, FDR-corrected $p=0.005$) at zero additional LLM calls,
significant on all three dimensions. AS shows a weaker composite effect ($d=+0.14$, raw $p=0.050$), significant on completeness and usefulness
but not accuracy ($d=+0.03$, $p=0.640$). We treat these four main
effects as the paper's planned comparisons and correct them jointly
with Benjamini--Hochberg FDR (Table~\ref{tab:maineffects}): CC and ToC
remain significant after correction, but AS's corrected $p=0.066$ does
\emph{not} clear the conventional 0.05 threshold. We therefore describe
AS below as a non-significant trend, not a confirmed effect, rather than
treating a raw $p=0.050$ as ``borderline significant.'' SE is null on every dimension ($d=-0.01$; per-dimension breakdown in
Table~\ref{tab:maineffects-full}). CC's
effect is more than double ToC's and more than triple AS's. The full
per-dimension breakdown is in Table~\ref{tab:maineffects-full}
(Appendix~\ref{sec:extra}); those per-dimension tests are secondary and
exploratory, reported as raw, uncorrected $p$-values, since the four
composite-level main effects are the only planned comparisons.

\begin{table}[t]
\centering
\small
\begin{tabular}{@{}lcccccc@{}}
\toprule
\textbf{Feature} & \textbf{On} & \textbf{Off} & \textbf{$\Delta$} & \textbf{$d$} & \textbf{$p$ (raw)} & \textbf{$p$ (FDR)} \\
\midrule
CC  & 4.24 & 3.85 & $+0.39$ & $+0.48$ & $<$0.001 & $<$0.001 \\
ToC & 4.13 & 3.95 & $+0.18$ & $+0.22$ & 0.003 & 0.005 \\
AS  & 4.10 & 3.98 & $+0.12$ & $+0.14$ & 0.050 & 0.066 \\
SE  & 4.04 & 4.04 & $-0.01$ & $-0.01$ & 0.894 & 0.894 \\
\bottomrule
\end{tabular}
\caption{Main effects on composite quality ($n=759$), with
Benjamini--Hochberg FDR correction across these four planned
comparisons. CC and ToC dominate and remain significant after
correction; AS does not survive correction; SE is null.}
\label{tab:maineffects}
\end{table}

\subsection{Feature Interactions}
\label{sec:interactions}

\begin{figure}[tbp]
\centering
\includegraphics[width=\columnwidth]{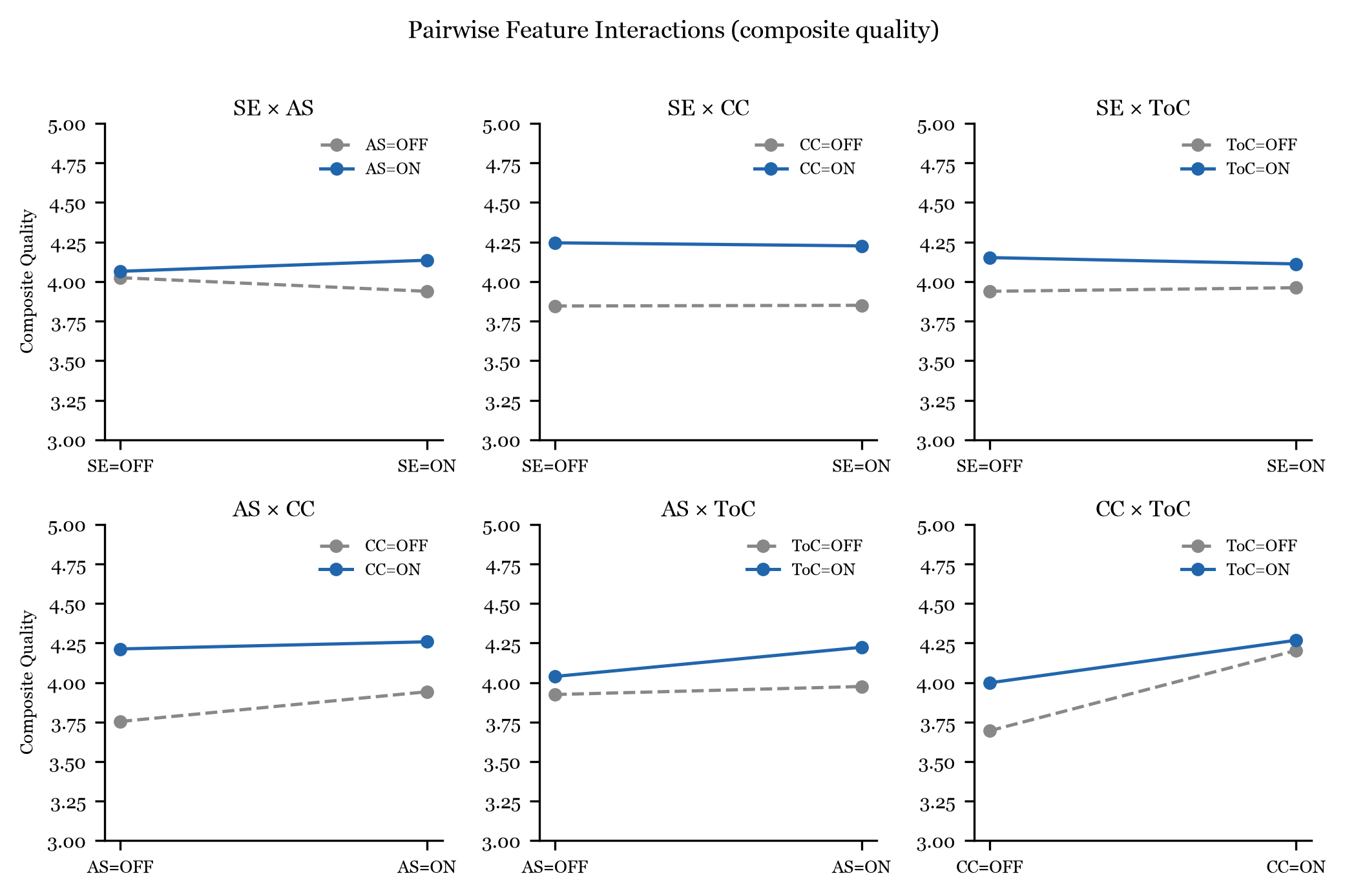}
\caption{Two-way interaction plots for all six feature pairs; each panel
shows composite quality as a function of one feature at both levels of the
other. CC dominates every panel it appears in; CC\,$\times$\,ToC shows the
largest interaction (the lines converge, so each feature adds less when
the other is on); SE adds nothing at either level of CC.}
\label{fig:interactions}
\end{figure}

A factorial design's distinctive payoff is detecting interactions that
one-at-a-time testing cannot see. Figure~\ref{fig:interactions} plots all
six two-way interactions. \textbf{CC dominates every panel it appears
in}: CC-ON quality exceeds CC-OFF at both levels of the other feature.
CC's benefit is not fully independent of retrieval, however. The
CC\,$\times$\,ToC panel holds the largest two-way interaction in the
design ($-0.12$, bootstrap $z=-2.04$, 95\% CI $[-0.23,-0.01]$): CC adds
$+0.51$ when ToC is off but $+0.27$ when it is on, and ToC adds $+0.30$
without CC but only about $+0.06$ with it, so the two features appear to
cover some of the same gaps. We treat this overlap as suggestive rather
than established. It is the only one of the eleven interaction terms
whose CI excludes zero, which is about what chance alone would produce
across eleven tests, and although its sign holds under all three
alternate scorings (\S\ref{sec:judgerobust}), it shrinks to $-0.07$
under both alternate LLM judges, with CIs that include zero. SE, by
contrast, is null whether CC is off ($\approx0.00$) or on ($-0.02$); the
SE\,$\times$\,CC panel shows two flat lines, so SE's lack of value is a
main effect, not an interaction. The remaining four pairs are modest (all
interaction terms within $\pm0.08$, $|z|\le1.4$). No pair shows a large
synergistic boost; the unequal main effects (CC large, SE null), together
with the possible CC/ToC overlap, are the mechanism behind the ``which
features, not how many'' result. One-at-a-time testing cannot see the
overlap: it would report ToC's isolated $+0.29$ and miss that ToC adds
only about $+0.06$ once CC is enabled.

\paragraph{Higher-order interactions.} The full $2^4$ design also
identifies three-way and four-way interactions, computed with the
standard factorial contrast (the signed sum of the 16 configuration means
divided by 8). All four three-way terms are small
(SE$\times$AS$\times$CC $+0.08$, SE$\times$AS$\times$ToC $-0.02$,
SE$\times$CC$\times$ToC $-0.04$, AS$\times$CC$\times$ToC $+0.05$; all
$|z|\le 1.4$) and the four-way term is smaller still ($-0.03$,
$z=-0.51$). All are at most about a fifth of CC's main effect ($+0.39$),
and none is distinguishable from zero. The CC\,$\times$\,ToC term above
is the one exception: at $-0.12$ it is the largest interaction term of
any order and the only one whose 95\% CI excludes zero. Apart from that
possible overlap, the factorial is well described by an additive model.

\subsection{Query-Type Moderation}

\begin{figure}[tbp]
\centering
\includegraphics[width=\columnwidth]{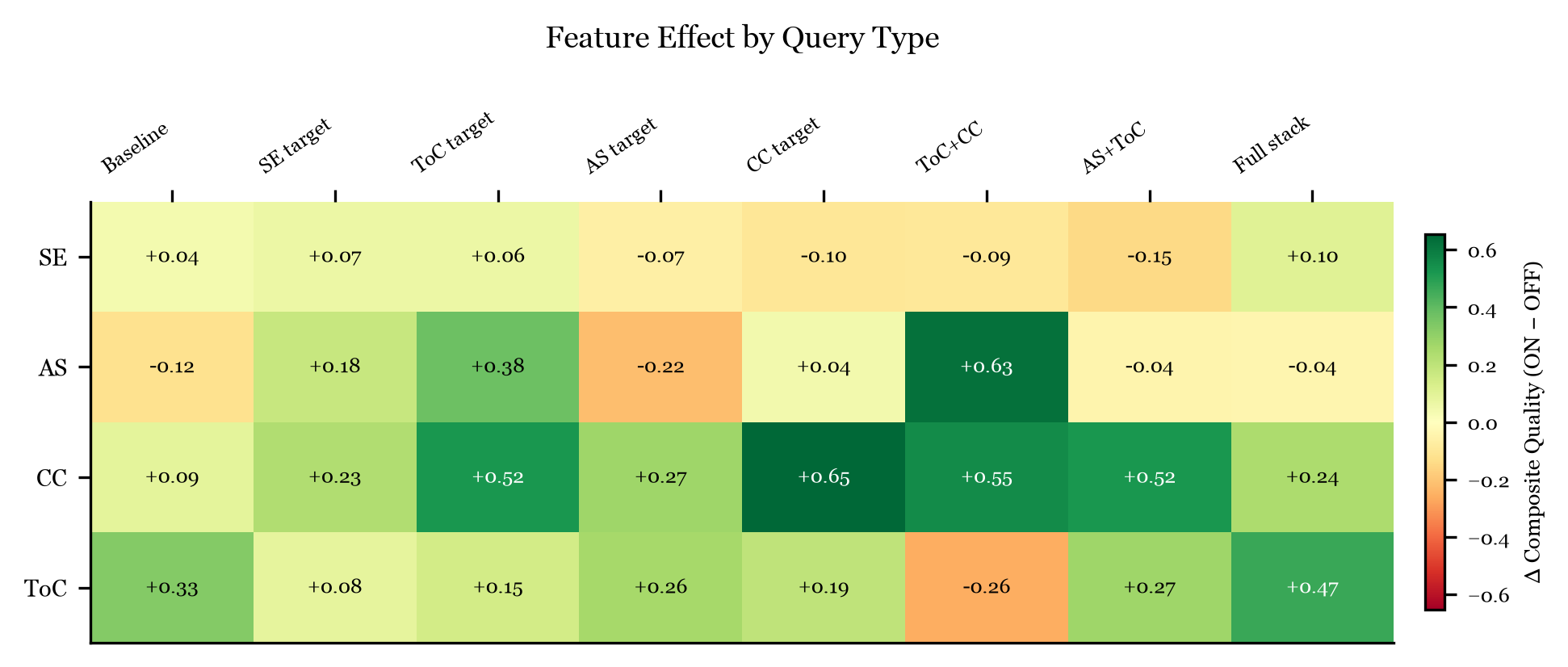}
\caption{Feature effect on composite quality ($\Delta=$ mean quality with
feature ON minus OFF) by query interaction type. Green: improvement; red:
degradation. CC is positive on every type; AS is strongest on \texttt{toc\_cc}
but negative on \texttt{as\_target}; ToC is negative on \texttt{toc\_cc}.}
\label{fig:heatmap}
\end{figure}

Feature utility is strongly moderated by query type
(Figure~\ref{fig:heatmap}), and aggregate effects mask substantial
variation. CC shows its strongest effects on \texttt{cc\_target}
($\Delta=+0.65$) and \texttt{toc\_cc} ($\Delta=+0.55$) queries,
confirming that completeness verification adds the most value where
exhaustive coverage is demanded; a targeted subgroup test gives
CC~$\rightarrow$~\texttt{cc\_target} completeness $d=+0.83$, $p<0.001$.
ToC is strongest on \texttt{full\_stack} ($+0.47$) and baseline
($+0.33$) queries. AS is the most unstable feature: it helps on
\texttt{toc\_cc} ($+0.63$) and \texttt{toc\_target} ($+0.38$) but is
\emph{negative} on \texttt{as\_target} ($\Delta=-0.22$; subgroup
completeness $d=-0.27$, $p=0.19$), suggesting that decomposition can
fragment retrieval rather than improve it; given that AS's aggregate
effect does not survive FDR correction (\S\ref{sec:main-effects}), these
subgroup splits should be read as directional and exploratory rather than
individually conclusive. The taxonomy's targets are only partly borne
out: CC peaks on its own target type, but AS is most negative on
\texttt{as\_target}, and ToC is negative on \texttt{toc\_cc} ($-0.26$), a
type designed to need it. An evaluation using only one
query type would systematically over- or under-value AS, and it would not
always confirm CC's advantage: ToC shows the larger gain on baseline
($+0.33$ vs.\ $+0.09$) and \texttt{full\_stack} ($+0.47$ vs.\ $+0.24$)
queries, and AS on \texttt{toc\_cc} ($+0.63$ vs.\ $+0.55$). Table~\ref{tab:querytype} gives
the absolute quality per type, which orders roughly as designed, from
baseline queries (4.71, easiest) down to \texttt{cc\_target}
exhaustive-listing queries (3.30, hardest). This difficulty gradient was
built into the taxonomy, however, and so is not itself independent
evidence; the more informative result is the feature-specific
\emph{directional} pattern above (e.g., AS helping on \texttt{toc\_cc}
but hurting on \texttt{as\_target}), which the design does not
guarantee in advance.

\begin{table}[tbp]
\centering
\small
\begin{tabular}{@{}lcccc@{}}
\toprule
\textbf{Type} & \textbf{$n$} & \textbf{Quality} & \textbf{Acc.} & \textbf{Compl.} \\
\midrule
baseline    & 95  & $4.71\pm0.55$ & 4.81 & 4.54 \\
full\_stack & 96  & $4.45\pm0.55$ & 4.44 & 4.27 \\
se\_target  & 92  & $4.40\pm0.51$ & 4.07 & 4.43 \\
as\_target  & 95  & $4.09\pm0.58$ & 3.88 & 3.93 \\
as\_toc     & 64  & $3.93\pm0.81$ & 3.97 & 3.77 \\
toc\_target & 127 & $3.89\pm0.86$ & 3.91 & 3.68 \\
toc\_cc     & 95  & $3.59\pm0.74$ & 3.84 & 3.36 \\
cc\_target  & 96  & $3.30\pm0.91$ & 3.20 & 3.07 \\
\bottomrule
\end{tabular}
\caption{Quality by query interaction type ($\pm$SD), ordered roughly as
designed by construction; the feature-specific directional findings
discussed above are the independent evidence this ordering alone does
not provide.}
\label{tab:querytype}
\end{table}

\subsection{Quality--Latency Pareto Frontier}

\begin{figure}[tbp]
\centering
\includegraphics[width=\columnwidth]{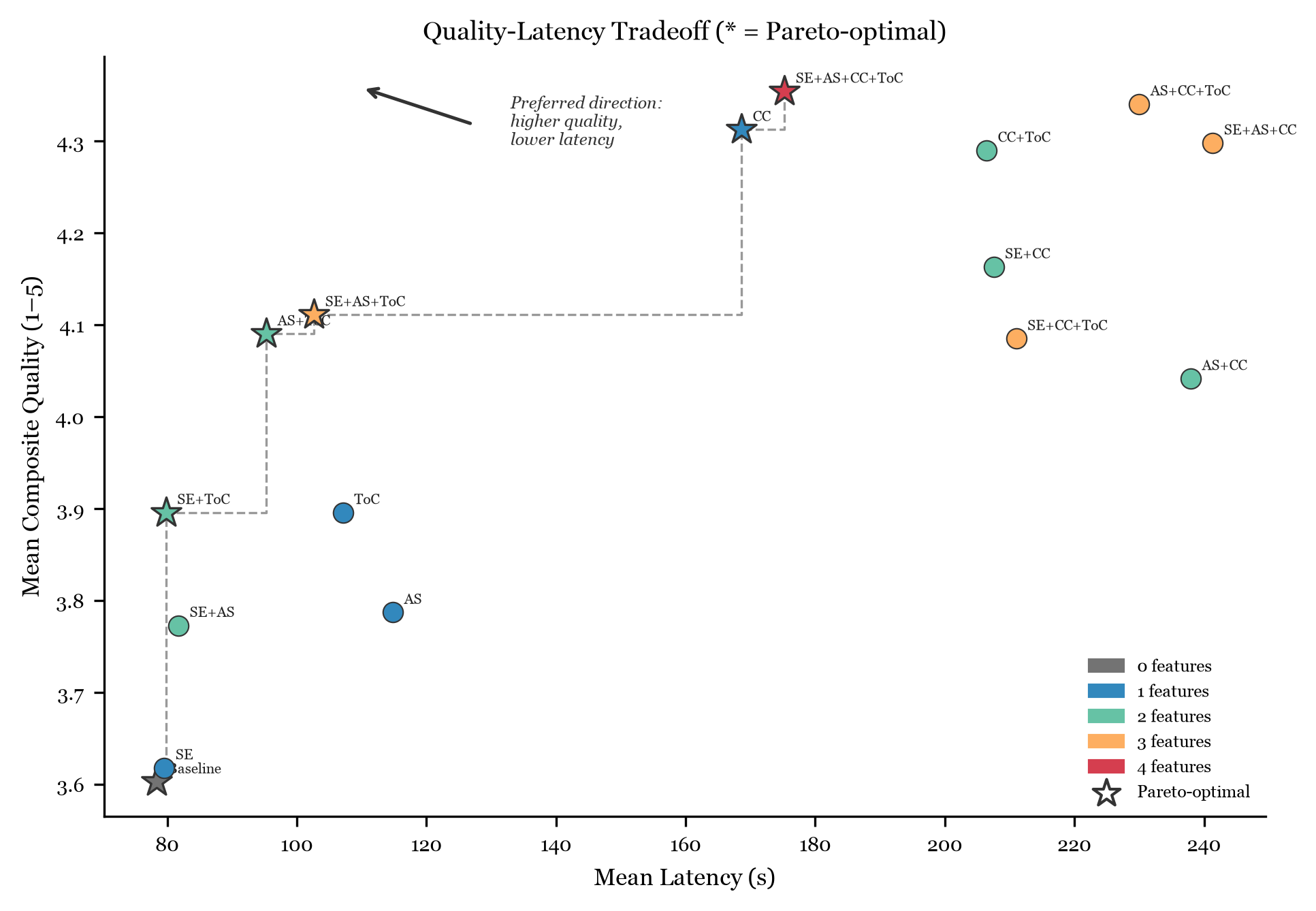}
\caption{Quality--latency frontier; each point is one configuration
averaged over both models and 24 queries. Color encodes feature count
(0--4); stars mark the six Pareto-optimal configurations in
Table~\ref{tab:pareto}, with latency compared at 1s resolution.
CC-containing configurations form the high-latency cluster.}
\label{fig:pareto}
\end{figure}

Figure~\ref{fig:pareto} plots all 16 configurations in quality--latency
space. Six configurations are Pareto-optimal (Table~\ref{tab:pareto}), forming a genuine
tradeoff: the all-features configuration is highest-quality (4.35) but at
175s, more than double baseline (78s); CC alone (4.31 at 169s) is a close
second. Below that, AS+ToC (4.09 at 95s) and SE+ToC (3.90 at 80s) offer
meaningful gains at near-baseline latency. The frontier shows a clear
latency step: CC-containing configurations occupy the high-latency region
(169--241s; quality 4.04--4.35), while non-CC configurations occupy the
lower-left (3.60--4.11 at 78--115s). The quality ranges overlap, but the
latency ranges do not. No
configuration reaches CC-level quality at non-CC latency. CC is the
primary latency driver: CC-containing configurations average 210s versus
92s without CC. We compare latencies at the whole-second resolution
reported in Table~\ref{tab:configs}, under which SE alone (3.62, 80s) is
dominated by SE+ToC (3.90, 80s). On unrounded means SE alone would join
the frontier, being 0.25s faster than SE+ToC, a gap far inside either
mean's roughly 7s standard error. More generally, the frontier is
computed on point estimates, and several adjacent members (e.g., AS+ToC
and SE+AS+ToC) differ by less than their standard errors.

\subsection{Retrieval-Level Impact and Robustness}
\label{sec:retrieval-robustness}

Because only ToC encodes page numbers in its evidence IDs, page-level
precision/recall isolate ToC's retrieval contribution. An independent
cross-check found the originally recorded ground-truth pages misaligned
with the actual source files for four of the five documents; we describe
the discovery and correction here in full, since it overturns a
conclusion drawn from the original ground truth.

\paragraph{How the error was found.} A second annotator, independent of
the original query author, re-derived relevant pages for all 24 queries
directly from the source PDFs in the public corpus, blind to the
original ground truth. For GDPR query V01 (``What is the definition of
personal data under GDPR?''), the original ground truth recorded page
33; the second annotator recorded page 3. Both cannot be reading the
same page content for the same question, so we checked the actual PDF:
page 3 of the corpus file contains Article 4's text verbatim
(``\,`personal data' means any information relating to an identified or
identifiable natural person''), while page 33 discusses an unrelated
provision on data protection impact assessment consultations. The
second annotator was correct. The same pattern held for UNOPS query V03
(``What are the four solicitation methods...''): the corpus PDF's page
62 contains the verbatim answer (``Shopping, request for quotation
(RFQ), invitation to bid (ITB) or request for proposal (RFP)''), while
the original ground truth's page 45 discusses procurement planning
levels, an unrelated topic. We verified this pattern by direct content
inspection, not just page-number comparison, for at least one query per
document across all five documents.

The misalignment affects GDPR, the UN Staff Rules, the UNOPS
Procurement Manual, and the EU AI Act: every document except NIST SP
800-53r5, whose original ground truth we independently confirmed
correct by the same method. This does not appear to be an annotation error: the pattern
is consistent with the corpus PDFs having been revised or re-sourced at
some point after the original ground truth was recorded, although we
did not log the exact date of the original annotation and so cannot
directly confirm the ordering from our own records. One document
supplies textual evidence consistent with this explanation: the UNOPS
Procurement
Manual's current text states in-place that a section ``has been amended
as per the Authoritative Revision to UNOPS Procurement Manual, Ref
2024/07,'' indicating that document has been revised at least once since
its original 2021 edition \citep{unopsprocurement2021}; we cite the
original edition's reference details because that is what we have on
file, while the PDF we re-checked is the amended text. We also note,
without asserting it as a
confirmed mechanism, a structural correlation: the four affected
documents all use layout-inferred heading extraction (\texttt{heading\_extraction})
in our ingestion pipeline, while the one unaffected document (NIST)
uses the PDF's native embedded table of contents (\texttt{native\_toc}).
We looked for a page-numbering bug in the extraction code that would
explain this and did not find one; in particular, the original
ground-truth pages were recorded by a human reading the source
documents directly, not derived from the pipeline's own page mapping, so
a bug confined to that code could not by itself explain the
discrepancy. We therefore report the \texttt{heading\_extraction}
correlation as an open question rather than a settled cause. The more
likely explanation is simply that the four affected source files were
updated to newer editions after annotation, consistent with the UNOPS
revision notice above. All ground-truth pages and all retrieval
metrics reported in this paper use the corrected values; the original,
since-superseded ground truth is kept in the public data release
alongside the correction for transparency. Because the public PDFs for
these documents can evidently change over time, we archive the exact
files used (not just links to them) in the same public data release.

\paragraph{Corrected retrieval metrics.} With corrected ground truth,
ToC pages hit relevant content for 15 of the 24 queries (62.5\%, versus
5 of 24, or 20.8\%, under the original ground truth); when they hit,
precision averages 0.56 (originally 0.38) and recall 0.32 (originally
0.15). Hit rate by query type
(Appendix~\ref{sec:extra}, Fig.~\ref{fig:retrieval}) is: \texttt{cc\_target}
and \texttt{as\_toc} 100\%, \texttt{toc\_target} 75\%, \texttt{as\_target}
and \texttt{full\_stack} 67\%, and baseline, \texttt{se\_target}, and
\texttt{toc\_cc} 33\% each. Several of these subgroups rest on as few
as two unique queries (Table~\ref{tab:querytypes}; see Limitations) and
should be read as descriptive rather than conclusive. This changes a specific claim: the
original, uncorrected analysis reported a clean 0\% hit rate on
baseline, \texttt{cc\_target}, and \texttt{se\_target} queries and
concluded that ``ToC fires on structural-retrieval needs rather than
indiscriminately.'' That conclusion does not survive the correction:
\texttt{cc\_target} is now the \emph{highest}-hit-rate query type, tied
with \texttt{as\_toc}, and baseline queries (designed as a null control)
hit relevant content a third of the time. ToC hits relevant content
across every query type to some degree; there is no clean type-level
separation between where it fires usefully and where it does not. This
correction is confined to the page-level retrieval metrics: ToC's
answer-quality main effect ($d=+0.22$, \S\ref{sec:main-effects}) is computed from
LLM-judge scores on full answers, independent of this page-matching
procedure, and is unaffected by it.

\paragraph{Reference provenance.} The same audit uncovered a second,
separate issue. The deployment's retrieval index also held a few internal
documents of the author's institution, most likely left over from an
earlier internal study, and two gold references (V05, on promotion
criteria, and V23, on purchase-order amendments) had been written from
those internal documents rather than from the public corpus. The public
PDFs do not contain these answers (we searched their text for the
distinctive content), which also explains why the second annotator rated
both queries low-confidence. Responses to three other queries (V01, V07,
V13) also cited internal documents. Because the main analysis should
reflect the experiment as run, we keep all 24 queries and report a
sensitivity check instead. Excluding V05 and V23 leaves the conclusions
unchanged: CC ($d=+0.51$) and ToC ($d=+0.22$) remain significant after
FDR correction, SE stays null ($d=-0.02$), and CC alone (4.33) still
exceeds every non-CC configuration (best 4.10). Excluding all five
affected queries gives the same picture (CC $d=+0.55$, ToC $d=+0.23$).
The exception is AS, which becomes significant without these queries
($d=+0.16$, FDR-corrected $p=0.0495$; $d=+0.20$, $p=0.020$ without all
five), so its non-significance on the full query set is fragile rather
than evidence of no effect. In the public data release, institution
names in model outputs that quote the internal documents are masked.

SE's null aggregate effect ($d=-0.01$) is consistent with two factors:
where heading hierarchies are deep, sibling chunks are less
query-relevant, and ToC already loads the same-section context SE
targets, at a coarser (page) granularity. Thirty-nine responses (5.1\%) score below 2.5;
these concentrate on three multi-part queries (32 of the 39) and are distributed
across both models and across configurations with and without features,
indicating query-level difficulty rather than
feature-induced degradation.

\begin{figure}[tbp]
\centering
\includegraphics[width=\columnwidth]{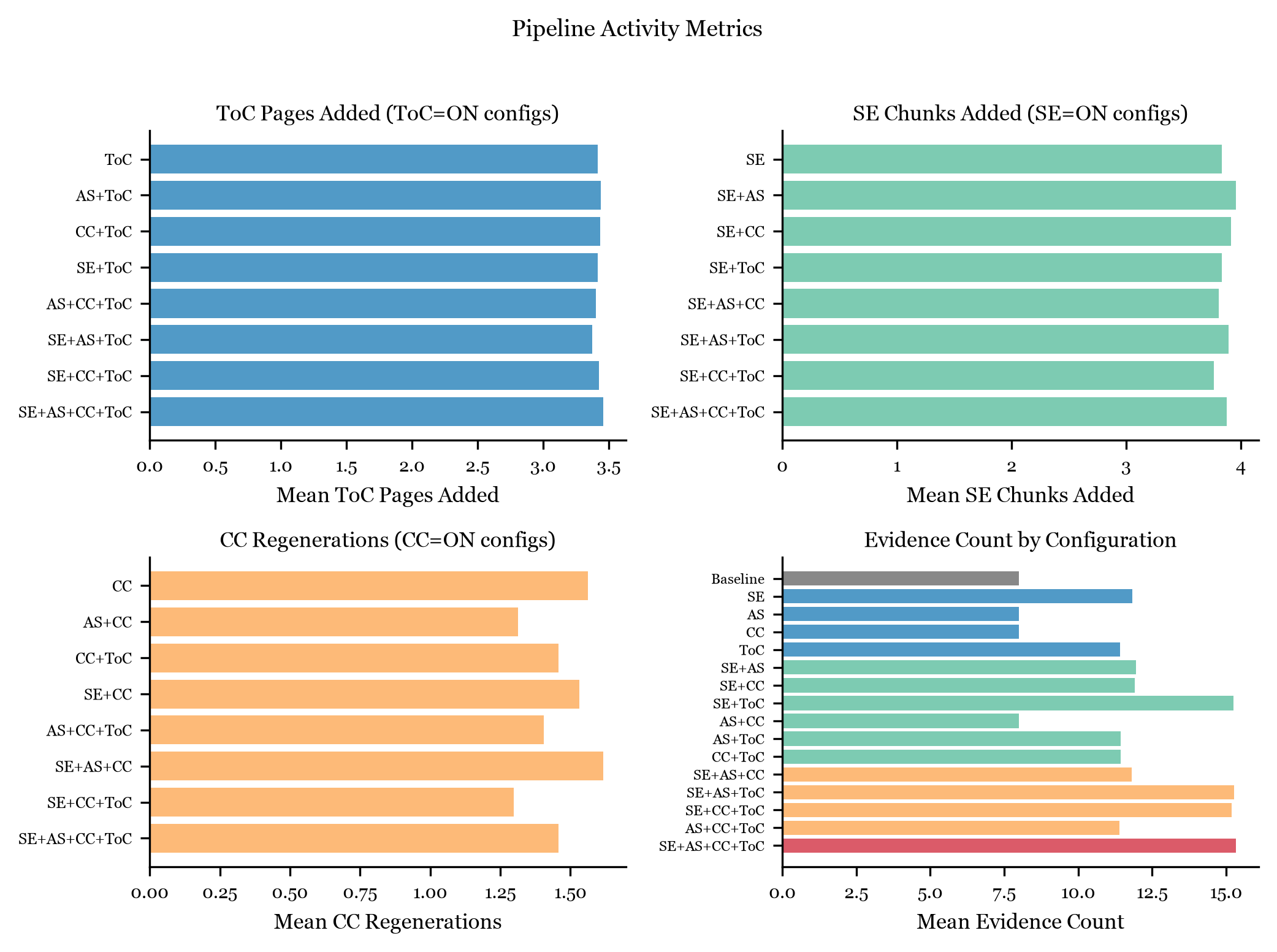}
\caption{Pipeline activity. (A) ToC pages added; (B) SE chunks added;
(C) CC regenerations; (D) total evidence items in each condition's final
evidence set by configuration: 8 base retrieval chunks (AS replaces
rather than adds chunks) plus SE and ToC additions. ToC and SE add
roughly constant evidence; CC tends to regenerate less when paired with
ToC.}
\label{fig:pipeline}
\end{figure}

\paragraph{How features change retrieval.}
Figure~\ref{fig:pipeline} traces what each feature does to the pipeline.
ToC adds $\sim$3.4 pages per query and SE adds $\sim$3.9 sibling chunks,
both roughly constant regardless of which other features are active. CC's
regeneration count, by contrast, shifts with ToC: adding ToC lowers it in
three of the four CC configuration pairs (mean 1.51 to 1.40 per query;
the exception is AS+CC, already low at 1.31), suggesting that proactive
page retrieval pre-covers some facets CC would otherwise flag as missing,
consistent with the possible CC/ToC overlap in \S\ref{sec:interactions}. Latency and accuracy--completeness breakdowns appear in
Appendix~\ref{sec:extra}.

\section{Discussion}
\label{sec:discussion}

\paragraph{Verification beats retrieval optimization.}
CC's mechanism (decompose the question into facets, check each against
the generated answer, re-retrieve for gaps) is fundamentally different
from the retrieval-stage features. SE, AS, and ToC all try to get better
evidence \emph{into} the context window; CC verifies whether the
generated answer actually \emph{used} that evidence to cover every
facet. On this corpus, query set, and implementation, post-generation
verification matters more than retrieval-stage optimization for answer
quality, but not for cost. CC's gains come at roughly double the
latency of the cheapest configurations (Table~\ref{tab:configs}), so
``dominates'' describes quality, not a cost-equivalent comparison; see
the Pareto frontier below for the tradeoff this implies. Notably, CC
improves accuracy as well as completeness
(Appendix~\ref{sec:extra}, Fig.~\ref{fig:acccomp}): its re-retrieval
cycle not only fills gaps but sharpens correctness.

\paragraph{A retrieval-vs-verification tradeoff.}
The Pareto frontier is better read as retrieval versus verification:
retrieval-stage features improve quality cheaply but with diminishing
returns, while verification achieves substantially larger gains at
substantial latency cost. The right operating point depends on the
deployment's latency budget. For quality-critical settings, enable CC by
default: it alone reaches 4.31, above any non-CC configuration, and its
169s is the price of completeness verification. For latency-sensitive
settings, ToC is the strongest single feature requiring zero additional
LLM calls ($+0.29$ over baseline), and AS+ToC (4.09 at 95s) is a strong
balanced choice. SE's aggregate effect is null on this corpus
(Table~\ref{tab:maineffects}), though the single-run latency figures in
Table~\ref{tab:configs} are noisy enough that SE's apparent effect on
latency specifically (e.g.\ ToC at 107s vs.\ SE+ToC at 80s, and the
175s-to-230s gap between the maximum-quality configuration and the same
configuration without SE) should not be read as a reliable cost, only
quality, comparison; we did not collect repeated-measurement latency
data that would let us separate real effects from run-to-run API
variance here. Table~\ref{tab:deployment}
summarizes the recommended configuration for each operating constraint.

\begin{table}[tbp]
\centering
\small
\begin{tabular}{@{}llcc@{}}
\toprule
\textbf{Scenario} & \textbf{Config} & \textbf{Qual.} & \textbf{Lat.} \\
\midrule
Minimum latency  & Baseline      & 3.60 & 78s  \\
Zero added LLM calls & ToC           & 3.90 & 107s \\
Balanced         & AS+ToC        & 4.09 & 95s  \\
Quality-focused  & CC            & 4.31 & 169s \\
Maximum quality  & SE+AS+CC+ToC  & 4.35 & 175s \\
\bottomrule
\end{tabular}
\caption{Recommended configuration by operating constraint. CC is the
common thread in the two high-quality options; ToC is the strongest
single feature requiring zero additional LLM calls. Latency values are
single-run figures (Table~\ref{tab:configs}); latency gaps involving SE
are not reliably estimated (see text).}
\label{tab:deployment}
\end{table}

\paragraph{Which features, not how many.}
CC alone (4.31) outperforms the three-feature SE+AS+ToC (4.11). Mean
quality rises with feature count, but variance within each count is
large, and the common thread in the highest-quality configurations is
CC, not feature count. Practitioners should prioritize \emph{which}
features to enable.

\paragraph{Methodological implications.}
Two design choices were decisive. \emph{Factorial design}: one-at-a-time
testing (each feature alone vs.\ baseline, from unrounded means) would
report SE $+0.02$, AS $+0.18$, CC $+0.71$, ToC $+0.29$, but only the
factorial shows that CC's gain falls from $+0.51$ to $+0.27$ when ToC is
on and that ToC adds only about $+0.06$ once CC is enabled, a possible
overlap that one-at-a-time testing cannot see. \emph{Query-type diversity}: CC's aggregate
gain ($\Delta=+0.39$) masks a range from $+0.09$ on baseline queries to
$+0.65$ on \texttt{cc\_target} queries (completeness $d=+0.83$), and AS,
ToC, and SE all flip sign across types. Evaluations using only simple
queries would systematically undervalue CC and AS and overvalue ToC.

\paragraph{Future work.}
Three directions follow. \emph{Dynamic CC triggering}: CC's latency comes
from firing on every query, including those the baseline already answers;
a query-complexity gate could restrict it to likely-incomplete answers.
\emph{Corpus-stratified ablation}: rerunning the 16 configurations across
structurally diverse corpora would identify which document properties
moderate each feature. \emph{Cost-aware feature selection}: a
meta-controller choosing features per query under a latency budget could
optimize the quality--cost frontier dynamically.

\section{Conclusion}

This $2^4$ factorial ablation (16 configurations $\times$ 24 queries
$\times$ 2 models; 768 runs, 759 scored) on five public
documents yields four findings, scoped to this corpus, task, and
implementation. (1) \textbf{Post-generation verification
dominates retrieval-stage optimization on quality, not cost}: CC
($d=+0.48$) is the single
most impactful feature, improving accuracy, completeness, and usefulness
simultaneously, and CC alone (4.31/5) is ahead of every configuration
lacking it (a descriptive comparison; see Table~\ref{tab:maineffects}
for the tested main effect), at roughly double the latency of the
cheapest configurations. Both results hold under both alternate LLM
judges.
(2) \textbf{Structure-aware retrieval is significant at zero additional
LLM calls}: ToC shows a significant main effect ($d=+0.22$,
FDR-corrected $p=0.005$) at zero additional LLM calls; it is also the only
feature whose page-level retrieval could be measured, and with corrected
ground truth it surfaces relevant pages on every query type. (3) \textbf{CC helps on every query type, but feature rankings are
query-dependent}: CC is positive on all eight types ($\Delta=+0.09$ to
$+0.65$) and the strongest feature on five, while ToC leads on baseline
and \texttt{full\_stack} queries and AS on \texttt{toc\_cc}. AS trends
toward helping some cross-section queries ($+0.63$) and harming others
($-0.22$), though AS's aggregate effect does not survive FDR correction
on the full query set (FDR-corrected $p=0.066$; it does once the two
internally sourced queries are removed) and these subgroup splits are
exploratory, not confirmed effects. (4) \textbf{Which features matter
more than how many}: CC alone beats the three-feature SE+AS+ToC, and the
six Pareto-optimal configurations span 0--4 features, with CC in both of
the highest-quality ones. Together these argue that verifying answers is a higher-leverage
investment than optimizing retrieval, and that factorial designs with
diverse query types are necessary to evaluate RAG features without
mis-ranking them.

\section*{Limitations}
\label{sec:limitations}

\textbf{Two models of similar capability.} Both answer models,
DeepSeek-V4-Flash and grok-4.3, are cloud-class; feature effects may
differ with local or smaller answer models (the AS and CC review model is
already a local 8B model). \textbf{Public regulatory documents only.} The corpus is
public regulatory/policy text (apart from the internal documents noted in
\S\ref{sec:retrieval-robustness}); effects on technical manuals,
code documentation, or financial reports may differ. \textbf{Retrieval
metrics limited to ToC.} Page-level precision/recall is measurable only
for ToC-sourced evidence, since other features operate at the chunk or
generation level without page tracking. \textbf{SE's null may be
corpus-specific.} SE's value likely depends on heading depth and
sibling-chunk relevance, which vary across corpora; shorter, shallower
documents may benefit. \textbf{Anchored-only evaluation.} Model-answer
anchored scoring cannot measure dimensions absent from the gold
standard, such as novel valid perspectives or user-specific relevance.
Scoring relies on a single primary LLM judge; \S\ref{sec:judgerobust}
(Table~\ref{tab:judgerobust}) reports five robustness checks against it:
two independent blind human raters, two alternate LLM judges (one
cross-provider and one a later same-family model), and a reworded-prompt
re-score of the same judge. All are consistent with a largely cancelable
calibration offset, with moderate rather than perfect agreement on
individual responses, and the four main effects keep their pattern under
every alternate scoring. The two human raters agree with each other
about as well as either agrees with the judge ($r=0.61$ between raters
vs.\ $r=0.69$ and $r=0.64$ individually against the judge), suggesting
part of the gap reflects genuine subjectivity in 1--5 answer-quality
judgments rather than a defect specific to this judge. These remain small-sample
checks (two human raters, not a properly powered multi-rater panel),
which remains future work. \textbf{Query ground truth, independently
cross-checked.} The 24 queries' relevant-page ground truth was
originally authored by a single annotator. A second, independent
annotator later re-derived relevant pages for all 24 queries directly
from the current source files, blind to the original answers. As
detailed in \S\ref{sec:retrieval-robustness}, this cross-check found the original pages misaligned
with the actual document files for four of the five source documents
(every document except NIST SP 800-53r5, independently confirmed
correct), and all ground-truth pages and retrieval metrics reported in
this paper use the corrected values. This correction still rests on a
single independent re-derivation per query rather than multiple
independent annotators, so some residual page-level uncertainty on
individual queries may remain; a fully independent re-annotation by a
third party, or a formal inter-annotator-agreement study, remains future
work. We built a custom query set in the first place because, to our
knowledge, no existing RAG/QA benchmark provides page-level,
verifiably-annotated ground truth over long, real-world regulatory and
policy documents with rich heading structure; this reflects our search
rather than an exhaustive survey. \textbf{Small per-type samples and
custom implementations.} Each query-interaction type is exercised by
only 2--4 unique queries (repeated across 16 configurations and 2
models); bootstrapped confidence intervals characterize sampling
variability in the scored responses but do not correct for this
underlying query-level clustering. Each feature is also a single,
custom implementation with specific engineering choices (Appendix~\ref{sec:settings}) rather than a standardized reference version, so
results characterize these implementations rather than the general
technique.
\label{end:body}

\bibliography{references}

\appendix

\section{Experiment Settings}
\label{sec:settings}

\paragraph{Retrieval and indexing.}
Embedding model nomic-embed-text (768-d, local Ollama); chunk size 512
tokens with 64 overlap; hybrid pgvector cosine + BM25 fused by weighted
reciprocal rank (weights 0.7/0.3, $k=60$); retrieval top-$k$ 8 with a
$4\times$ candidate multiplier (32 vector and 32 BM25 candidates fused to
8); source diversity enabled; 8{,}192-token context window. With these
weights, any vector candidate outscores any BM25-only candidate, so BM25
can only reorder chunks both channels find; it did so for 2 of the 384
baseline evidence items.

\paragraph{Feature settings.}
SE: min 2 siblings, max 15 chunks. AS: forced LLM decomposition on every
query (production heuristic gate disabled), max 4 sub-queries, original
query embedding preserved; its merged evidence held 8 chunks in every
run, so AS changes which chunks are retrieved, not how many. CC: max 2 hops,
max 3 sub-queries, max 15 evidence items, facet-decomposition prompt,
runs independently of ToC. ToC: page budget 5, match top-$k$ 3, min
similarity 0.30, auto mode.

\paragraph{Models and scoring.}
Answer models DeepSeek-V4-Flash and grok-4.3 (Azure AI). Review model
(CC facet checker and AS decomposer) llama3.1:8b. Scorer
Claude Opus 4.6 (Azure, direct API), scoring accuracy/completeness/%
usefulness (1--5) at temperature 0, max 500 tokens; composite is their
unweighted mean.

\paragraph{Statistics.}
Welch's $t$-test for main effects; Cohen's $d$ with pooled SD; bootstrap
CIs with 5{,}000 resamples (seed 42, 95\% percentile interval); standard
error $\text{SD}/\sqrt{n}$.

\section{Additional Results}
\label{sec:extra}

\begin{figure}[t]
\centering
\includegraphics[width=\columnwidth]{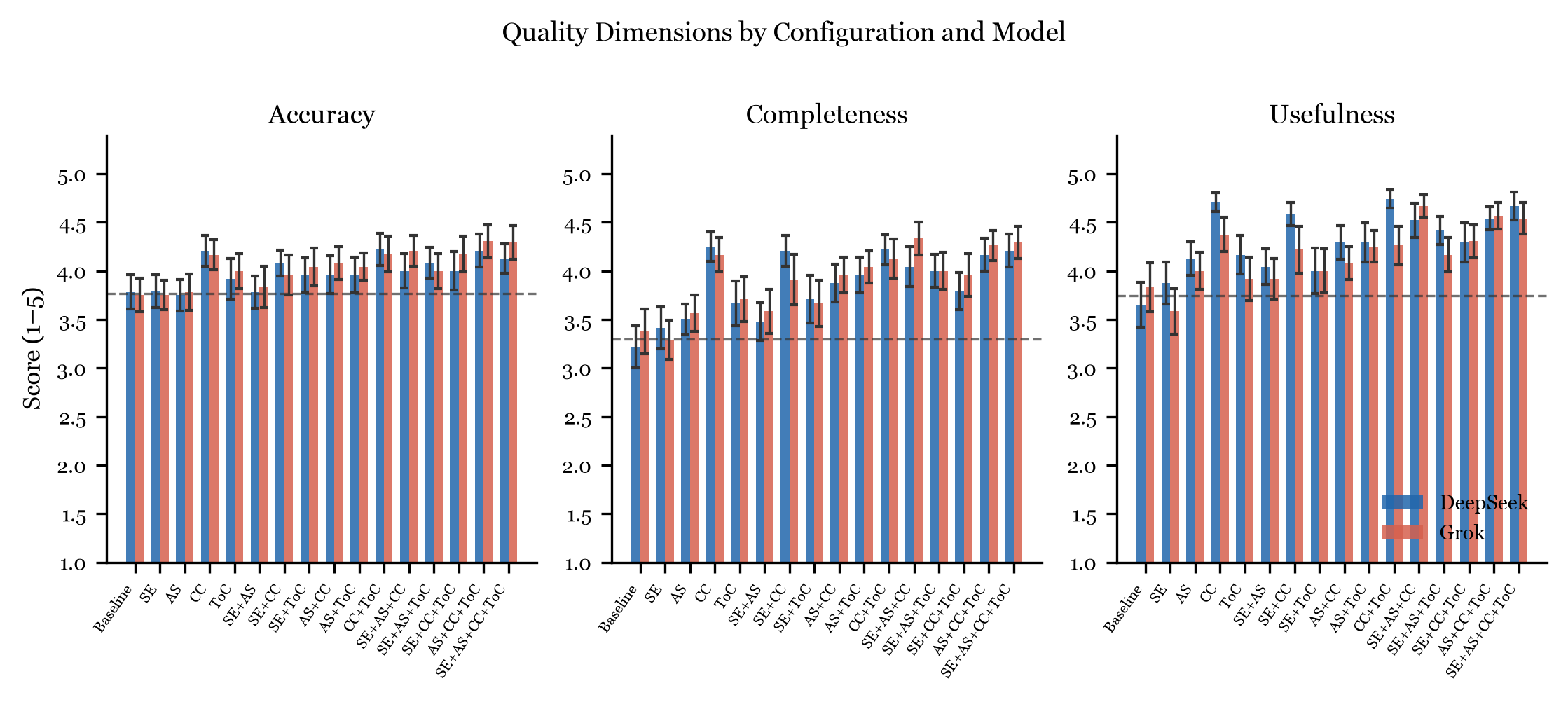}
\caption{Accuracy, completeness, and usefulness (1--5) for each of 16
configurations, grouped by model. Error bars: SEM; dashed line: baseline
mean per dimension.}
\label{fig:dims}
\end{figure}

\begin{figure}[t]
\centering
\includegraphics[width=\columnwidth]{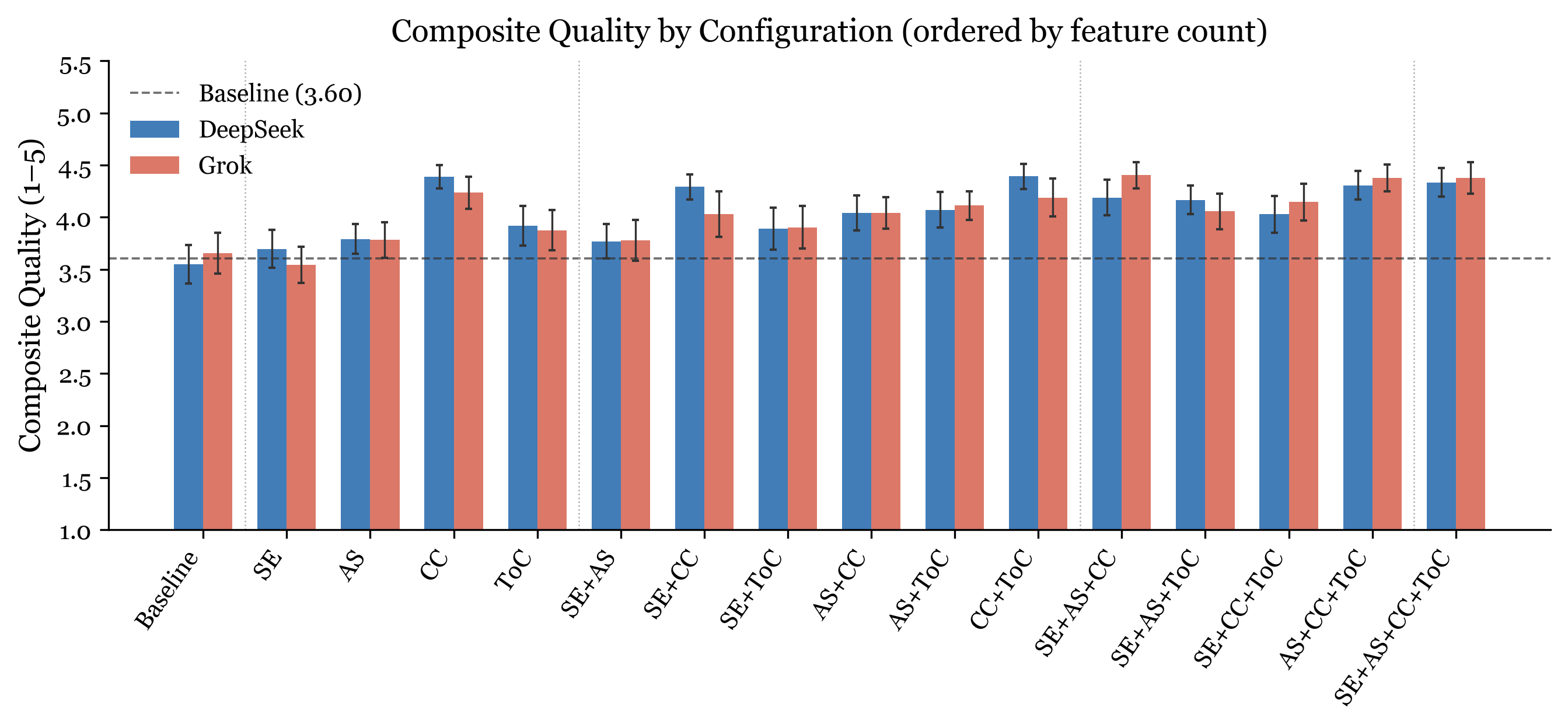}
\caption{Composite quality by configuration and model, ordered by feature
count. Dashed line: baseline composite (3.60). CC alone exceeds all
non-CC configurations.}
\label{fig:composite}
\end{figure}

\begin{figure}[t]
\centering
\includegraphics[width=\columnwidth]{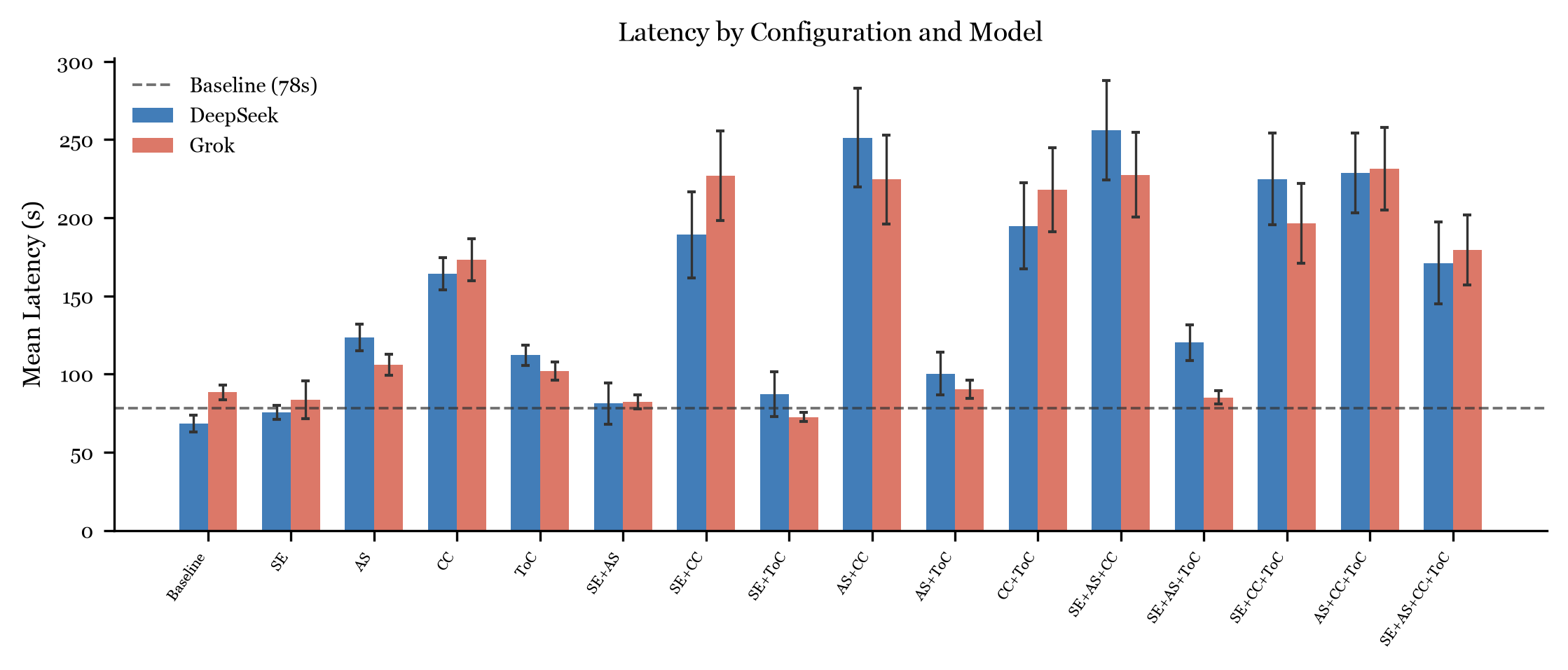}
\caption{Mean latency by configuration and model (error bars: SEM; dashed
line: baseline). CC is the primary latency driver.}
\label{fig:latency}
\end{figure}

\begin{figure}[t]
\centering
\includegraphics[width=\columnwidth]{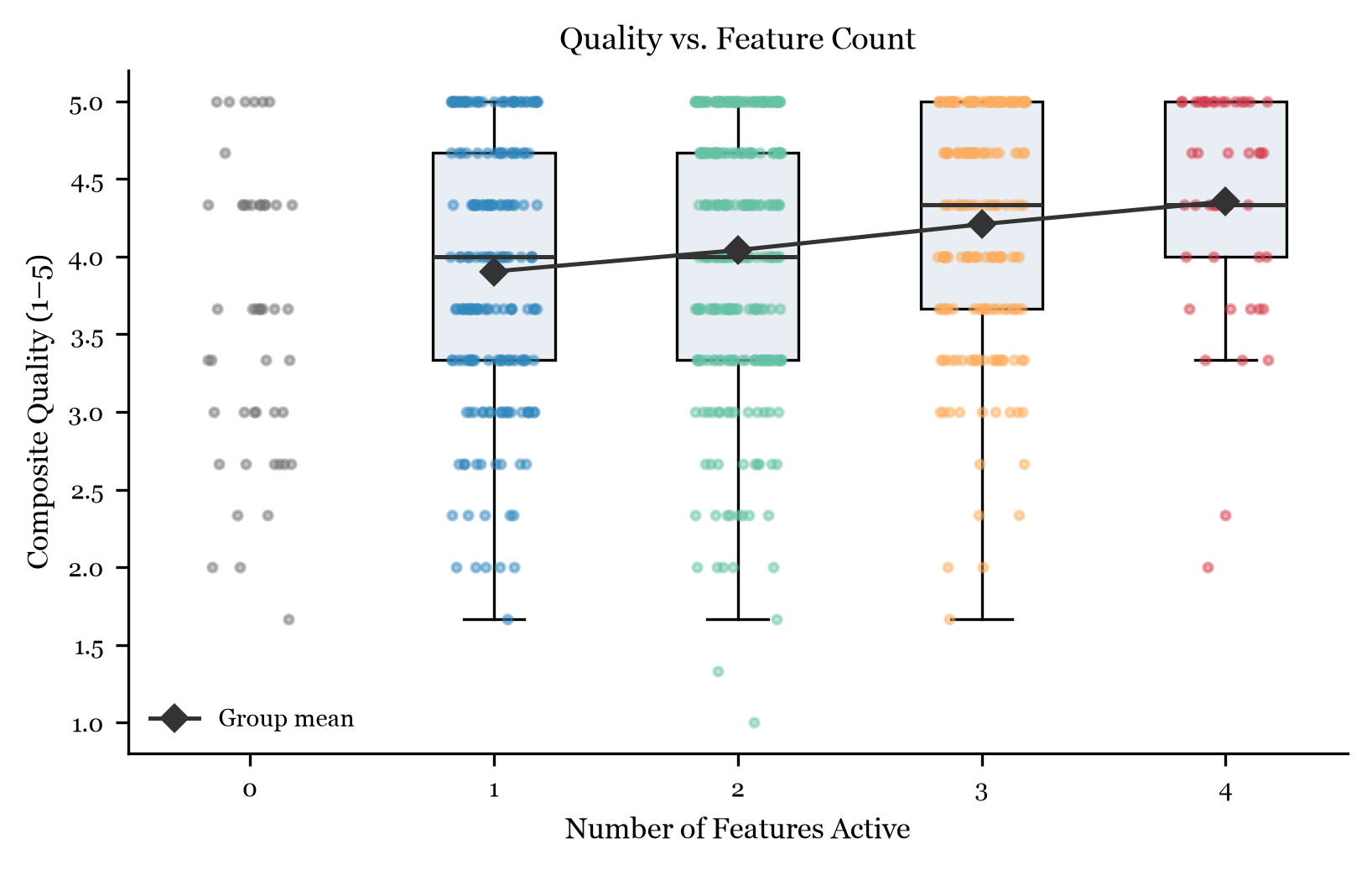}
\caption{Composite quality by number of enabled features. Means (diamonds)
rise with feature count, but within-group variance is large, especially
at 1--2 features where CC-containing configurations dominate.}
\label{fig:nfeatures}
\end{figure}

\begin{figure}[t]
\centering
\includegraphics[width=\columnwidth]{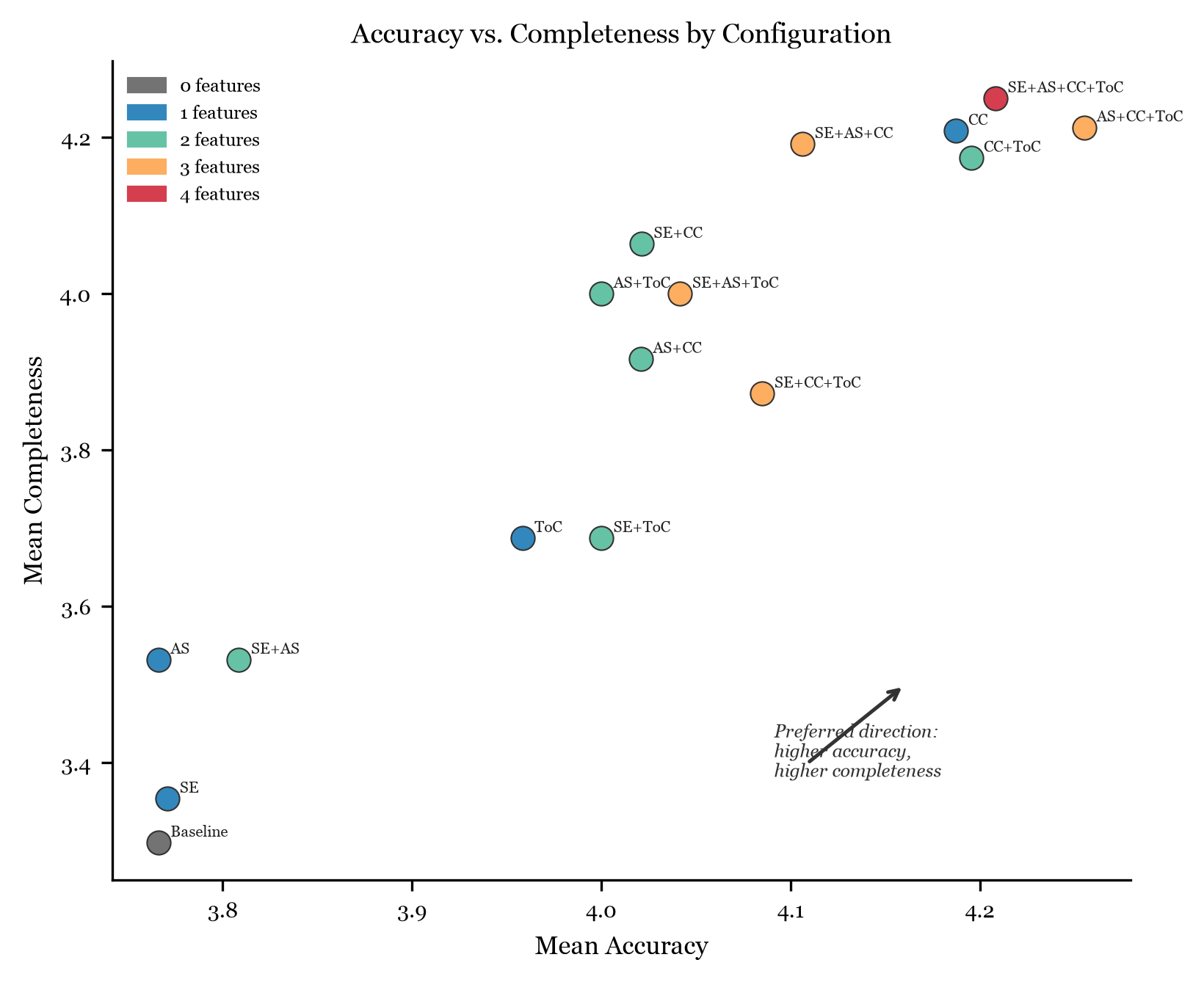}
\caption{Mean accuracy vs.\ completeness per configuration (color: feature
count). Most CC-containing configurations cluster upper-right, improving both
dimensions simultaneously.}
\label{fig:acccomp}
\end{figure}

\begin{figure}[t]
\centering
\includegraphics[width=\columnwidth]{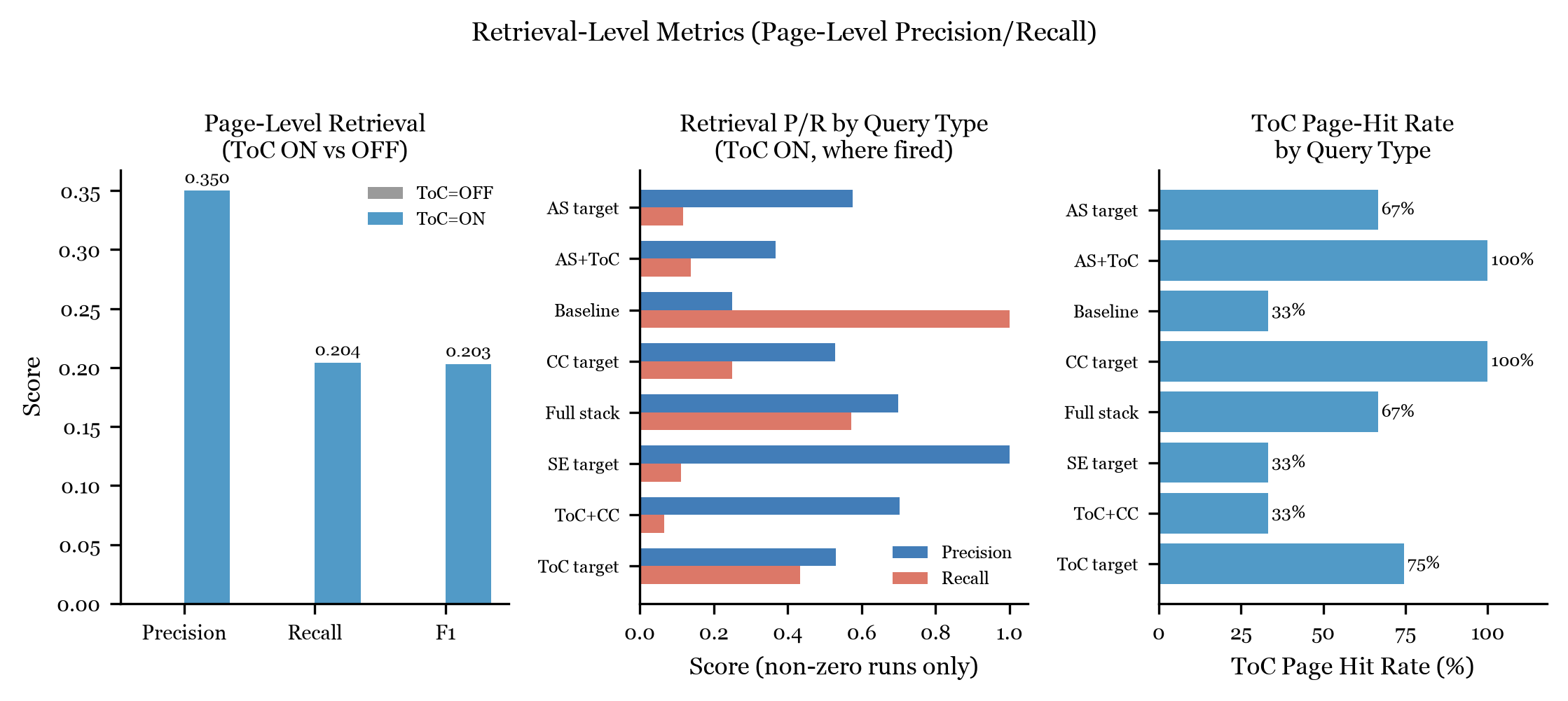}
\caption{Page-level retrieval metrics, computed from the corrected
ground truth. Left: precision/recall/F1 averaged over all
ToC-ON runs, including misses (ToC-OFF runs carry no page-level evidence,
so those bars are empty by construction). Center: precision/recall by query type on ToC-hit runs. Right:
ToC page-hit rate by query type.}
\label{fig:retrieval}
\end{figure}

Table~\ref{tab:maineffects-full} gives the full per-dimension main
effects and Table~\ref{tab:pareto} lists the Pareto-optimal
configurations. The remaining figures provide the per-dimension
(Fig.~\ref{fig:dims}), per-configuration (Fig.~\ref{fig:composite}),
latency (Fig.~\ref{fig:latency}), feature-count (Fig.~\ref{fig:nfeatures}),
accuracy--completeness (Fig.~\ref{fig:acccomp}), and page-level retrieval
(Fig.~\ref{fig:retrieval}) breakdowns.

\begin{table}[t]
\centering
\small
\begin{tabular}{@{}llccccc@{}}
\toprule
\textbf{Feat.} & \textbf{Dim.} & \textbf{On} & \textbf{Off} & \textbf{$\Delta$} & \textbf{$d$} & \textbf{$p$} \\
\midrule
CC  & Qual. & 4.24 & 3.85 & $+0.39$ & $+0.48$ & $<$0.001 \\
    & Acc.  & 4.13 & 3.89 & $+0.25$ & $+0.29$ & $<$0.001 \\
    & Compl.& 4.11 & 3.64 & $+0.47$ & $+0.49$ & $<$0.001 \\
    & Use.  & 4.46 & 4.02 & $+0.44$ & $+0.49$ & $<$0.001 \\
ToC & Qual. & 4.13 & 3.95 & $+0.18$ & $+0.22$ & 0.003 \\
    & Acc.  & 4.09 & 3.93 & $+0.16$ & $+0.19$ & 0.009 \\
    & Compl.& 3.98 & 3.76 & $+0.22$ & $+0.23$ & 0.002 \\
    & Use.  & 4.32 & 4.16 & $+0.16$ & $+0.18$ & 0.016 \\
AS  & Qual. & 4.10 & 3.98 & $+0.12$ & $+0.14$ & 0.050 \\
    & Acc.  & 4.03 & 4.00 & $+0.03$ & $+0.03$ & 0.640 \\
    & Compl.& 3.96 & 3.79 & $+0.16$ & $+0.17$ & 0.022 \\
    & Use.  & 4.32 & 4.16 & $+0.16$ & $+0.18$ & 0.016 \\
SE  & Qual. & 4.04 & 4.04 & $-0.01$ & $-0.01$ & 0.894 \\
    & Acc.  & 4.01 & 4.02 & $-0.01$ & $-0.02$ & 0.831 \\
    & Compl.& 3.87 & 3.88 & $-0.01$ & $-0.01$ & 0.887 \\
    & Use.  & 4.24 & 4.24 & $-0.00$ & $-0.00$ & 0.993 \\
\bottomrule
\end{tabular}
\caption{Main effects by quality dimension ($n=759$). Qual.\ $=$
composite quality; Compl.\ $=$ completeness.}
\label{tab:maineffects-full}
\end{table}

\begin{table}[t]
\centering
\small
\begin{tabular}{@{}lccc@{}}
\toprule
\textbf{Config} & \textbf{Quality} & \textbf{Latency} & \textbf{Pareto} \\
\midrule
SE+AS+CC+ToC & 4.35 & 175s & yes \\
CC           & 4.31 & 169s & yes \\
SE+AS+ToC    & 4.11 & 103s & yes \\
AS+ToC       & 4.09 & 95s  & yes \\
SE+ToC       & 3.90 & 80s  & yes \\
Baseline     & 3.60 & 78s  & yes \\
\bottomrule
\end{tabular}
\caption{Pareto-optimal configurations, with latency compared at the 1s
resolution shown. SE alone (3.62, 80s) is dominated by SE+ToC (3.90,
80s); on unrounded means it is 0.25s faster and would also qualify.}
\label{tab:pareto}
\end{table}

\section{Scoring Rubric and Reproducibility}
\label{sec:repro}

Each completed response was scored by Claude Opus 4.6 against a verified
gold-standard reference for its query, called directly via Azure (not
through the pipeline) to avoid retrieval contamination. The scorer
received the question, the reference answer (used as the rubric), and the
candidate, and scored \emph{accuracy} (facts correct vs.\ reference),
\emph{completeness} (fraction of reference content covered), and
\emph{usefulness} (actionability) on 1--5. Reference answers were
generated from the source documents (except V05 and V23;
\S\ref{sec:retrieval-robustness}) and manually reviewed for factual
accuracy before the experiment. The scoring prompt, condensed here
(verbatim, with its JSON output request, in the archived
\texttt{run\_v3\_experiment.py}), was:

\begin{quote}\small
You are evaluating a knowledge assistant's answer against a gold-standard
reference answer. Be strict and objective.\\[2pt]
\textbf{Question:} \texttt{\{query\}}\\
\textbf{Reference} (all expected facts; use as the rubric):
\texttt{\{model\_answer\}}\\
\textbf{Candidate to evaluate:} \texttt{\{answer\}}\\[2pt]
Score each dimension (1--5) against the reference. \emph{Accuracy}: are
the candidate's facts correct versus the reference? (1 $=$ multiple
errors, 3 $=$ mostly correct, 5 $=$ precise). \emph{Completeness}: what
fraction of the reference's content is covered? (1 $=$ misses most, 3 $=$
majority, 5 $=$ all key points). \emph{Usefulness}: would a practitioner
find it actionable? (1 $=$ vague, 3 $=$ adequate, 5 $=$ clear, structured).
\end{quote}

The 24 queries span GDPR (5), UN Staff Rules (5), EU AI Act (5), UNOPS
Procurement (5), and NIST 800-53 (4), each with ground-truth relevant
page numbers, relevant articles, predicted retrieval behavior, and target
hypotheses. The runner streams pipeline events over SSE, caches each
condition's raw JSON for resumable execution, retries failed queries up
to twice (300s timeout) and scoring up to three times with exponential
backoff. All 11 figures are generated at 300 DPI from a single analysis
CSV with fixed seeds (bootstrap seed 42). The raw data (760 completed
responses, 759 scored, with full telemetry; institution names in outputs
that quote the internal documents are masked), the 24 query definitions
with references and both the original and corrected ground-truth pages,
the alternate-judge and human-rater scores, scripts that reproduce the
robustness checks, and the exact source-document files used (since, per
\S\ref{sec:retrieval-robustness}, the public versions of four of them
have since changed) are archived on Zenodo at
\url{https://doi.org/10.5281/zenodo.20986976}.

\section{Query Set}
\label{sec:queries}

Table~\ref{tab:queryset} lists all 24 queries with their interaction type,
source document, and ground-truth relevant pages, corrected per
\S\ref{sec:retrieval-robustness}.

\begin{table*}[t]
\centering
\scriptsize
\setlength{\tabcolsep}{4pt}
\begin{tabular}{@{}llcp{7.2cm}p{2.6cm}@{}}
\toprule
\textbf{ID} & \textbf{Type} & \textbf{Doc} & \textbf{Query} & \textbf{Pages} \\
\midrule
V01 & baseline & G & What is the definition of personal data under GDPR? & 3 \\
V02 & baseline & U & What is the normal retirement age for UN staff? & 68, 69 \\
V03 & baseline & P & What are the four solicitation methods available in UNOPS procurement? & 62 \\
V04 & se\_target & G & What are all the data subject rights under GDPR and how are they exercised? & 9, 12--22, 24, 30, 34, 68--70 \\
V05$^\dagger$ & se\_target & U & What are the eligibility criteria and time-in-grade requirements for promotion at each staff level? & 34, 37--39, 77 \\
V06 & se\_target & P & What are the different evaluation methodologies for each solicitation method in UNOPS procurement? & 101, 102 \\
V07 & toc\_target & G & What are the requirements for data breach notification: who must be notified, within what timeframe, and what information must be provided? & 29, 30 \\
V08 & toc\_target & U & What are the rules for annual leave accrual, carry-forward limits, and home leave eligibility for UN staff? & 35, 36, 42--46 \\
V09 & toc\_target & A & How does the EU AI Act classify a system as high-risk, and what use cases in Annex III qualify? & 53, 54, 124, 125, 127--129 \\
V10 & toc\_target & N & What access control policies does NIST 800-53 define, and how do they relate to identification and authentication? & 45--50, 63, 158--160, 163--166 \\
V11 & as\_target & U & Under what conditions can a UN staff member's appointment be terminated, and what indemnities are payable in each case? & 66, 69--71, 77, 78, 86, 87, 93, 94 \\
V12 & as\_target & A & What transparency obligations does the EU AI Act impose on providers and deployers, and how do they differ across high-risk, general-purpose, and human-interacting systems? & 58--60, 68, 69, 81--85, 100, 110, 130, 131, 136, 137, 141--143 \\
V13 & as\_target & N & What controls does NIST 800-53 recommend for protecting audit logs (generation, review, analysis, storage)? & 93--95, 97--104 \\
V14 & cc\_target & G & What are the conditions for lawful processing of personal data under GDPR and what makes consent valid? & 4, 7--11 \\
V15 & cc\_target & A & What penalties and fines does the EU AI Act prescribe, and how are they calculated for different infringements? & 115--118 \\
V16 & cc\_target & P & What are the rules for emergency and exigency procurement in UNOPS and how do they differ from standard procurement? & 62, 71, 72, 84, 86, 89, 174--179 \\
V17 & toc\_cc & U & Explain all the ways a UN staff member can be separated from service, including notice periods and financial entitlements for each. & 30, 53, 59--62, 66, 68--75, 77, 86, 87, 93--95, 110--113 \\
V18 & toc\_cc & A & What obligations must providers of high-risk AI systems fulfil before and after placing systems on the market? & 47, 49, 51, 55--65, 67, 77--83, 101, 102, 104--109, 130--136 \\
V19 & toc\_cc & P & Walk through the complete UNOPS procurement process from requirements definition to contract award, including review/approval thresholds at each stage. & 30, 32--35, 46, 48, 50, 51, 54, 57, 58, 60--63, 67, 71, 81, 89, 92, 94, 95, 99--102, 104, 109--111, 114--116, 120, 121 \\
V20 & as\_toc & A & What are the requirements for the risk management system under the EU AI Act, and how do they interact with conformity assessment for high-risk systems? & 47, 55--60, 62, 63, 76--78, 80, 101, 130, 131, 133--135 \\
V21 & as\_toc & N & How does NIST 800-53 address supply chain risk, and what controls from other families support it? & 246, 247, 279--282, 391, 392, 396 \\
V22 & full\_stack & G & What are all the requirements for appointing a Data Protection Officer under GDPR (when required, qualifications, role)? & 33--35 \\
V23$^\dagger$ & full\_stack & P & What are the different types of PO amendments, the steps for each, and how do they interact with PO closure and budget rules? & 32, 33, 51, 116, 120, 129, 132, 160--162 \\
V24 & full\_stack & N & What is NIST 800-53's complete incident-response framework, from planning and detection through recovery and lessons learned, and how does it integrate with contingency planning? & 142--147, 179--186 \\
\bottomrule
\end{tabular}
\caption{Complete query set with interaction type, source document, and
corrected ground-truth relevant pages (\S\ref{sec:retrieval-robustness}).
Doc codes: G $=$ GDPR, U $=$ UN Staff Rules,
A $=$ EU AI Act, P $=$ UNOPS Procurement Manual, N $=$ NIST SP 800-53r5.
$^\dagger$Reference answer drawn from internal documents rather than the
listed public document (\S\ref{sec:retrieval-robustness}).}
\label{tab:queryset}
\end{table*}

\end{document}